\documentclass[preprint,10pt,5p]{elsarticle}

\usepackage{graphicx}
\usepackage{float}
\usepackage{dcolumn}
\usepackage{bm}
\usepackage{amssymb}
\usepackage{multirow}
\usepackage{caption}

\journal{Physica A}

\begin{document}

\begin{frontmatter}

\title{Link deletion in directed complex networks}

\author[label1]{G. Kashyap}
\author[label1,label2]{G. Ambika\corref{cor1}}
\address[label1]{Indian Institute of Science Education and Research(IISER) Pune, Pune - 411008, India}
\address[label2]{Indian Institute of Science Education and Research(IISER) Tirupati, Tirupati - 517507, India}

\cortext[cor1]{Corresponding Author}

\ead[url]{http://www.iiserpune.ac.in/~g.ambika/}

\ead{g.ambika@iiserpune.ac.in}


\begin{abstract}
We present a systematic and detailed study of the robustness of directed networks under random and targeted removal of links. We work with a set of network models of random and scale free type, generated with specific features of clustering and assortativity. Various strategies like random deletion of links, or deletions based on betweenness centrality and degrees of source and target nodes, are used to breakdown the networks. The robustness of the networks to the sustained loss of links is studied in terms of the sizes of the connected components and the inverse path lengths. The effects of clustering and 2-node degree correlations, on the robustness to attack, are also explored. We provide specific illustrations of our study on three real-world networks constructed from protein-protein interactions and from transport data.
\end{abstract}


\end{frontmatter}


\section{Introduction}
\label{sec1}
The field of complex systems has caught the interest of the academic community and the industry in a big way in the recent past. It deals with how locally interacting constituents give rise to macroscopic structures and emergent phenomena. The study of these inter-dependencies is most effective with a complex network based approach. A comprehensive study of these networks would involve a long list of topics including descriptive analysis, study of generative models, dynamical or functional properties and more \cite{newman2003structure,strogatz2001exploring,boccaletti2006complex,albert2002statistical}. In this work, we are interested in studying their robustness. It is one of the more important aspects of a complex network and is the ability of the network to withstand and (or) adapt to internal and (or) external changes. While these changes can occur in many different ways, those resulting in loss of connectivity or breakdown of the network are of particular interest. In the not-so-recent past, we have been witness to multiple events that have persistently brought up the question of robustness. Be it global internet-failures, large-scale email attacks \cite{zou2004email}, country-wide electricity blackouts \cite{crucitti2004model,pagani2013power,carreras2002critical,dobson2007complex,carreras2004complex} or international stock-market crashes \cite{peron2011collective,kaue2012structure,may2008complex,chi2010network}, they have led us to ask the question: \textit{When does the failure of individual constituents translate to a failure of the whole?}

The connectivity of a network can be compromised as a result of random or targeted loss of nodes or edges. While the phenomenon of node-deletion has been extensively studied and used \cite{albert2000error,crucitti2003efficiency,cohen2000resilience,cohen2001breakdown,sole2008robustness,motter2002cascade,wang2009cascade,buldyrev2010catastrophic,crucitti2004error,santiago2009robustness,berche2010public,kawamoto2015precise,iyer2013attack}, to a large extent in undirected networks and to a lesser extent in directed networks, the avenue of link deletion has remained rather unexplored \cite{martin2006random,wang2009edge,he2009effect,goerdt2003analysis,zhang2007enhancing,callaway2000network,cartledge2011connectivity,holme2002attack}. Even the existing studies, concerning the robustness of directed networks\cite{fang2014modeling,yan2009random}, are limited to very specific contexts or systems and the utility of the results outside the respective contexts is minimal. There is a large number of real world networks, like biological, technological and transport networks, that are directed in nature. But there is no exhaustive study that covers the robustness of standard directed-network models to common methods of loss of links, in a systematic manner. Hence we find this area to be promising and worth exploring.

Link-deletion may seem similar to the process of node deletion, except that the node is still functional but unable to contribute to the overall functioning of the network. This perception changes once we realize that it is a much more fine-tuned way to manipulate the connectivity of a network. Also it leaves the node functional, which makes reconnection a possibility. Hence this process can be used to enhance the efficiency of flow on a network \cite{zhang2007enhancing} and also to constrain the flow if need be. For example, in a power grid, losing a few transmission lines, due to natural or man‐made reasons, could result in loss of power-supply. Besides lowering the efficiency of the grid, this could also destabilize other nodes or lines, due to overloading, thus leading to a cascading failure. Detaching the unstable nodes from the network by intentional and targeted link deletion, thus preventing a cascade of failures, can salvage this situation. Along similar lines, applications can also be seen in computer networks (to control the spread of a virus), disease spreading networks (prevent interactions to control the spread of a disease) and transport networks (how will a damaged road/track or a traffic jam effect the movement of traffic?). Another important application is in the study of some neuro-degenerative disorders like Alzheimer’s disease, where the loss of connectivity between neurons leads to lower efficiency in transmission of signals. Link deletion can also be used to optimize information flow on a network. Apart from the above, knowledge of the behavior of a network, when it is under different types of breakdown, helps us to identify patterns of structural and functional vulnerabilities and hence design robust systems that are resilient to various kinds of attacks and also put in place, precautionary measures that would come into effect in the event of breakdown.

In this paper, we present the results of our investigations on the effects of loss of links in directed complex networks with commonly observed properties like scale-free nature, clustering and degree-correlations. We intend to explore the structural changes occurring in the networks, in the event of failures (random loss of links) and attacks (targeted loss of links). To make this study systematic and more general, we work initially with a set of network models rather than actual real networks. In doing so, we also isolate the response of networks with specific properties to various kinds of link deletion mechanisms. We do a comparative study of the behaviors of various models of directed networks under different kinds of link deletion. In the later part of the paper, we present results of various types of link deletion on data drawn from real-world networks and look at possible interpretations of the same.

The paper is organized  as follows: In section II, we discuss the generation of directed networks with requisite properties like clustering and degree correlations and introduce the sample datasets that we have considered. In section III, we present the different strategies by which links are deleted. In this section, we also compare strategies based on local and global information and define the "edge-degree" in directed networks. In the following section, we identify a set of relevant measures that capture the changes occurring  in the network during link-deletion. In Section V we present the results of each link deletion strategy on different types of networks including three real world networks with specific properties. In section VI, we present the results followed by conclusions.


\section{Generation of directed networks}
\label{sec2}
Despite the wide spectrum of phenomena to which a network based approach can be directly applied, a bulk of the real-world networks exhibit some predominantly common characteristics. Thus many real networks exhibit a heavy-tailed degree distribution, leading to the famous power-law in the degree-distributions and consequent scale-free behaviour. Many such networks also exhibit the property of transitivity or clustering, to varying extents. Online Social Networks (OSNs) like Facebook, Twitter, Quora etc. have high values of clustering coefficients. Other examples of highly clustered networks include human or animal brain networks, friendship networks and co-authorship networks. But most technological networks and some biological networks like protein interaction networks, gene regulatory networks and metabolic networks show very little clustering. Another important property seen in many real-networks is the correlation between nodes, based on scalar or enumerative properties. For example, most social networks have been observed with high positive correlation (Assortativity) with respect to scalar characteristics like age, sex, hobbies etc. On the other hand, networks like the World Wide Web (WWW) and the internet show negative correlations (Dissortativity). Some other properties of networks include the existence of communities and hierarchical community structure and long-range correlations. 

With networks showing such a wide-array of properties, it becomes very tedious and challenging to both, model these networks and to study their response to internal failures and external attacks. Given this complexity in the networks, conducting a systematic analysis of their responses requires us to isolate the effects of loss of links on individual properties. Alternatively this also provides insight into how individual properties affect each other and the overall robustness of the network. In order to do this, we require ensembles of networks that dominantly exhibit only the relevant properties while others become statistically negligible. To generate such ensembles, we rely on  models of random and scale-free networks and also a set of degree preserving rewiring (DPR) mechanisms to introduce properties like 2-node degree-correlations and clustering into the generated networks. Below, we provide a brief description of the models used and introduce the real-world networks. 

\subsection{Directed Erd{\"o}s-R{\'e}nyi (ER) networks}
\label{ss21}
The ER model is among the very first to be introduced to model networks and is probably the most extensively studied one \cite{erdds1959random}. In this model, the network $G(N,M)$ starts with $N$ disconnected nodes. With uniform probability, $M$ directed edges are placed between these nodes such that no multi-edges and self-loops are introduced. This is among the most simplistic models with no inherent structural biases. The in and out degree distributions are Poissonian, with average degrees, $<k^{in}>$ and $<k^{out}>$, both equal to $M/N$. The characteristic path length of the network scales as the logarithm of the network size $N$ and the networks show negligible clustering and degree-correlations. For the present study, we generate ensembles of ER networks with 2500 nodes and average degree equal to 5 and subject them to various strategies of link-deletion.

\subsection{Directed Scale-Free (SF) networks}
\label{ss22}
Despite the importance of the ER model in deriving many analytic results concerning networks, it has its own share of shortcomings and therefore has little scope for application to real networks. One of the primary drawbacks of the ER model is the absence of heavy-tail in the degree distributions. This feature, along with the resulting scale-free behaviour, is a consistent observation in a majority of real networks. To study the effect of this scale-free behaviour on the robustness to loss of links, we employ the model proposed by Bollob{\'a}s et al. \cite{bollobas2003directed}. This model is based on a growth mechanism and generates directed networks with power-laws in both the in and out degree distributions. In the large $N$ limit, there are negligible degree-correlations and the clustering tends to zero. This allows us to study, in isolation, the effects of scale-free nature. Using this model, we generate ensembles of SF networks with 2500 nodes and $\sim$9000 links and in and out degree distributions with exponents equal to 2.5.

\subsection{Degree-correlated ER and SF networks}
\label{ss23}
As mentioned earlier, many real networks exhibit the property of 2-node degree-correlations. The correlations could be positive, as seen in social networks or negative as seen in technological and computer networks. To investigate their contributions, if any, to the robustness of the networks, we look at their role, both in isolation and in the presence of scale-free behaviour. To study their role in isolation, we generate random networks using the ER model and then introduce the desired type of correlation using the degree preserving rewiring mechanisms we have introduced recently \cite{doi:10.1093/comnet/cnx057}. Also, together with scale-free nature, we first generate the SF network using the model by Bollob{\'a}s et al. and then rewire the network to tune correlations to the required value. 

Owing to the direction associated with the links, 4 different types of 2-node degree-correlations are possible, namely In-In, In-Out, Out-In and Out-Out. Ensembles of ER/SF networks are generated with $N=2500$ and all 4 types of correlations are introduced approximately to the same extent ($\rho_q^p\approx\pm$0.4). The value of correlation is calculated using Spearman's rank correlation as given below.

\begin{equation}
\label{eqn:1}
\rho_q^p=\frac{12\sum_e R_e^pR_e^q-3M(M+1)^2}{M^3-M}
\end{equation} 

where $R_e^p$ and $R_e^q$ are the ranks given to source and target nodes associated with edge e, based on their p and q-degrees $(p,q\in\{in,out\})$, respectively, and $\rho_q^p$ is referred to as Spearman's Rho.

\subsection{Clustered ER and SF networks}
\label{sec:level24}
Another important property that is not captured in ER model or SF model \cite{bollobas2003directed}, but is observed in many real networks (social networks in particular), is clustering. Clustering is the ability of a network to form well-knit neighbourhoods and is quantified using the Mean Clustering Coefficient (MCC). The MCC is defined as the average of Local Clustering Coefficients (LCCs) of all nodes in the network. The LCC of a node is defined as the ratio of number of pairs of connected neighbours of the node (closed triplets or triangles) to the number of pairs of neighbours of the node (open triplets). Unlike in undirected networks, where there is no ambiguity in the type of triangles, in directed networks, 4 different types of directed simple closed triplets are possible: Cycles, Middleman triangles (Mids), In-triangles (Intri) and Out-triangles (Outri). 

We are interested in studying the role of clustering on the robustness of networks during loss of links. We first study the role of clustering in the absence of scale-free nature. This is done using the Watts-Strogatz (WS) model of small-world networks \cite{watts1998collective}, adapted to the case of directed networks. We start with an initial 1-D lattice containing N (=1000) nodes. Each node is connected to 3 neighbours with 3 in-links and 3 out-links. This gives a network with an average degree of 6. The directed links are then rewired with a rewiring probability $p_r=0.1$, resulting in a network with clustering coefficient $\sim0.4$ and in and out degree distributions that are close to poissonian. An ensemble of these networks is then subject to different strategies of deletion and the results are analyzed. To study the role of clustering in the presence of scale-free behaviour, SF networks are generated as mentioned earlier and the rewiring schemes introduced in \cite{doi:10.1093/comnet/cnx057} are used to tune the different MCCs to a desired value of $\sim0.2$).

\begin{figure*}[ht]
\centering
\includegraphics[width=\textwidth]{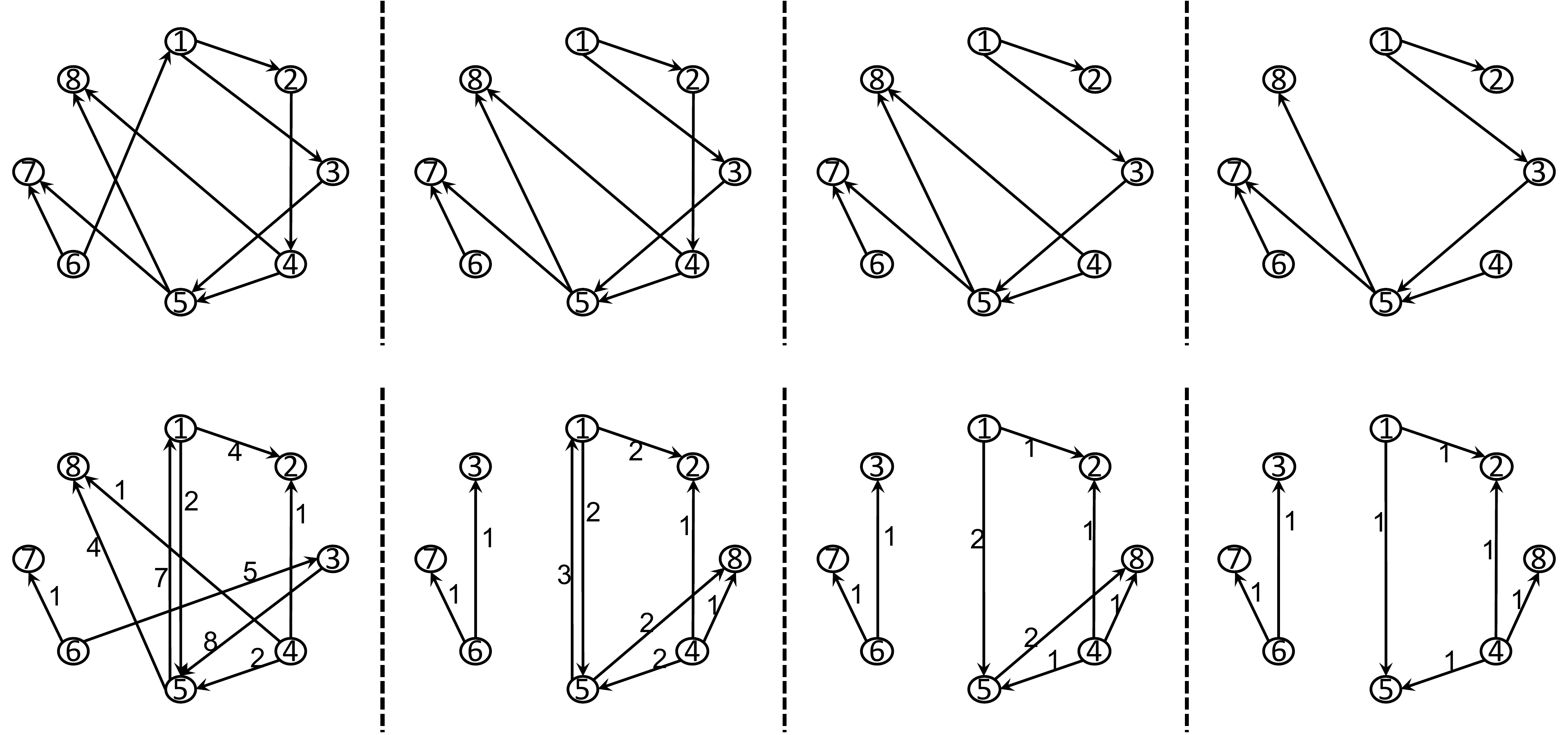}
\caption{A schematic depiction of the first few steps of link deletion (Left to Right). In the case of random deletion (Top), all edges have uniform weight and are equally likely to be deleted at any step. Therefore, the edge-weights have not been specified. In the case of targeted deletion (Bottom), the edge-weights, in this schematic, are specified by the EBC values of the edges but could in general be specified by any other metric. The edge with the highest weight is deleted at every step. If multiple edges carry the same weight, the tie is broken uniformly randomly. After every deletion, the network is allowed to reorganize and the values of EBC are modified accordingly and the process repeats.}
\label{fig:Demo}
\end{figure*}

\subsection{\label{sec:level25}Real-world Networks}
So far, we have discussed network models that generate specific desired properties. But these models need not always represent real networks. However by studying the effects of link deletion in these models, we get an unhindered view of the role of a specific property or set of properties in the response of networks to loss of links. This knowledge can be used to identify and analyze relevant results in real networks where many related and/or isolated properties can coexist. With this objective, we now introduce some real-networks on which we have studied the effects of link deletion. These datasets have been chosen to highlight the role of properties like SF nature, degree-correlations and clustering. A brief description for each dataset is given below. 

\textbf{\textit{Austin Road Network}}: As the first example we consider a road transport network, that represents the flow of traffic between road intersections in the city of Austin, TX. An intersection of two or more roads is represented as a node and if there is a flow of traffic between two intersections, then that is considered as a link. The dataset is sourced from \cite{github:2017:Austin} and although the original dataset contains detailed metadata, we will concern ourselves only with the edge-list of the network. This network consists of 7388 intersections and 18956 connections between them. The distribution of degrees is mostly homogeneous and Poisson-like. The network is highly assortative with respect to all types of 2-node degree-correlations but shows negligible clustering. As can be expected of city-roads, the connected components span the entire network. Being a transport network, it lends itself as an extremely relevant system for the study of link deletion. Many problems concerning the network, like shutting down of certain roads due to emergencies, unavailability of roads due to traffic jams or closure for maintenance purposes, can all be modelled as problems of link deletion. Besides, this particular network does not exhibit many other attributes and therefore we can study the singular effect of assortative connectivity in the network.

\textbf{\textit{Protein Interaction Network}}: This dataset comprises a network of interacting human proteins and was put together by U. Stelzl et al. \cite{konect:stelzl,konect:2017:maayan-Stelzl}, in an effort to better understand the organization of the human proteome. The network consists of 1706 proteins (nodes) undergoing 6171 unique interactions (links) as identified by high-throughput Y2H (Yeast Two-Hybrid) screening. Both the in and out degree distributions show scale-free behavior and there is a strong 1-node correlation. The network has negligible clustering and the strongly connected component spans 1493 nodes. The network is also considerably dissortative with respect to all 4 types of 2-node degree-correlations. Therefore, this dataset is a good example of a SF dissortative network on which we can study the effects of link deletion. Loss of links in this network can be equated to failed interactions between proteins. This could happen for a variety of reasons including random failures, structural mismatch, prohibitive media etc.. We are interested in exploring the robustness of this set of interacting proteins from the point of view of networks and wherever possible give a biological interpretation. 

\textbf{\textit{Airport Network}}: This network is constructed from data obtained from the openflights.org dataset\cite{konect:2017:openflights}. The nodes in the network represent various airports across the world and a link exists between two nodes if there is atleast one flight from the source to the destination airport. The network consists of 3425 airports connected by 37594 flights. The original network accounts for multiple flights between the same pair of airports and therefore has 67663 links and is a multigraph. For our work, we replace multiple edges by a single edge and generate a simple graph. As is expected, this network also shows very high 1-node correlation and the largest strongly connected component spans 3354 nodes. The network has very little 2-node degree-correlations but very large clustering coefficients with respect to all 4 types of triangles. For this reason, we have selected this network as a good representation of a clustered SF network. This network is subjected to different types of link deletion and its performance is analyzed with the help of the knowledge obtained from studying the network models. Studying link deletion on this network can help us to model the effects of cancellation of targeted flights on select routes on the dynamics of passenger traffic and the extent of connectivity.

The characteristic properties of all the networks presented above 
are summarized in Table-\ref{Tab:NetProp}. The values for degree correlations are averaged over the different types of possible 2-node correlations and the values of clustering coefficients are averaged over the values of clustering w.r.t various types of triplets.

\section{Strategies for link deletion}
\label{sec3}
When it comes to designing a strategy for deleting links, the list of options runs long and is limited mostly by the purpose of deletion and the type/amount of prior information required to implement such a strategy. We choose strategies so as to utilize, independently, both local and global information about the network. We assume that all necessary information is available which may not always be the case in some real-world scenarios. For example, in many cases, our knowledge of the network is incomplete and therefore we cannot gauge the full importance of a particular node/link. This could be because the network is too large or it could be a shortcoming in the data collection methods. In any case, we are forced to implement strategies with incomplete information. It could also happen that certain strategies require the evaluation of global metrics that are not always computationally viable. An immediate example is the computation of global properties, like shortest paths and betweenness centralities, which is very intensive and not practical for large networks containing millions of nodes/links. 
In this study, we consider both random and targeted deletion of links. In the case of targeted deletion, the links are assigned weights based on Edge-Degree(local-information) or Edge Betweenness Centrality (global-information). During targeted deletion, removal based on increasing and decreasing orders of weight are considered separately. Each strategy is discussed below in further detail. We also explore the relationship between Edge-Degree(ED) and Edge Betweenness Centrality(EBC) and motivate the importance of identifying such a relationship.

\subsection{Random Deletion}
\label{ss31}
In many real-world networks, loss of links occurs randomly and is unpredictable. This can be seen in loss of transmission lines in a power-grid due to inclement weather or loss of communication between routers on the internet due to wear and tear of the connecting cables. In the brain, random but progressive loss of synapses leads to devastating neuro-degenerative disorders like Alzheimer's Disease. Another common example is the cancellation of flights due to unforeseen circumstances that impacts the connectivity of airports around the world. The unpredictable nature of this type of loss can be correctly modelled as a uniform random removal of links in the network. At every step, a randomly chosen link is removed from the network (Fig.\ref{fig:Demo} - Top Panel) and the structural integrity of the network is measured as a function of loss of links. 
We refer to this type of random deletion of links as strategy S1.

\subsection{Deletion based on Edge Betweenness Centrality}
\label{ss32}
The betweenness centrality for links is a direct extension of the corresponding definition for nodes. Let $n_{ij}^e$ be the number of shortest paths between nodes i and j, that pass through link $e$ and $n_{ij}$ be the total number of shortest paths between i and j. Then, the EBC of $e$ is defined as the ratio of $n_{ij}^e$ to $n_{ij}$, summed over all possible pairs of nodes in the network. But at this stage, we are left with a stand-alone number, with limited utility. Therefore, this quantity is normalized by the total number of vertex pairs, so that EBC takes a value between 0 and 1.

\begin{equation}\label{eqn:2}
EBC_e=\frac{1}{N^2}\sum_{i,j}\frac{n_{ij}^e}{n_{ij}}
\end{equation}

In strategy S2, all the links are ordered in descending order of their EBC values and the link with the highest value of EBC is deleted at every step (Fig.\ref{fig:Demo} - Bottom Panel). In any network, the links with the highest EBC values are the ones that play the role of $'bridges'$ and connect groups of nodes that would otherwise be disconnected. Hence, these links play a very central role in ensuring connectivity between the two groups. The loss of these links would result in disconnected groups of nodes and hence absence of many shortest paths. Deletion based on centrality has been studied earlier in undirected networks and has been shown to be extremely effective in dismantling a network. 
After every instance of deletion, the network is allowed to reorganize and the EBC values are recalculated. Other network properties like degree-correlations, clustering coefficients and measures of robustness are also calculated at every step. 

\subsection{Deletion based on Edge-Degree}
\label{ss33}
While the notion of degree of a node seems quite natural, the degree of an edge is not equally intuitive. The concept of Edge-Degree was introduced for undirected networks, as the product of degrees of nodes on either sides of a link\cite{holme2002attack}. In this work, we modify this definition to suit directed networks, to be the product of in-degree of the source-node and out-degree of the target-node. The choice of this definition is justified in the later part of this section. If $k_i^{in}$ and $k_j^{out}$ are in and out degrees of source-node i and target-node j, respectively, then ED is given by (\ref{eqn:3}).

\begin{equation}
\label{eqn:3}
ED_e=k_i^{in}.k_j^{out}
\end{equation}

The motivation for defining ED is to find a quantity, based on which, weights can be assigned to links, using only local information about the links/nodes. Therefore, with (\ref{eqn:3}) as the working definition of ED, we assign weights to the links and order them in decreasing order of their weights. 

In strategy S3, the link with the highest value of ED is deleted, following which, the network is allowed to reorganize and the values of ED and other network properties are recalculated. Following the definition of ED, S3 targets the links that connect the  in-hubs to the out-hubs. Therefore, we expect the network to break down into multiple strongly connected components. 


\begin{table}[ht]
\centering
\caption{The values of correlation coefficient ($\rho$), between EBC and ED, for the various networks under consideration.}
\label{Tab:CorrCoeff}
\begin{tabular*}{0.8\columnwidth}{@{\extracolsep{\fill}} c c } 
\hline
\multirow{2}{*}{Network} & \multirow{2}{*}{$\rho (EBC,ED)$} \\ \\
\hline
Erd{\"o}s R{\'e}nyi (ER) & 0.840 \\
Scale-Free (SF) & 0.828 \\
Assortative ER & 0.613 \\
Dissortative ER & 0.847 \\
Assortative SF & 0.638 \\
Dissortative SF & 0.833 \\
Watts-Strogatz & 0.562 \\
Clustered SF & 0.717 \\
\hline
\end{tabular*}
\end{table}

\subsection{Relationship between Edge Betweenness Centrality and Edge Degree}
\label{ss34}
As mentioned earlier in this section, the key factors involved in designing effective strategies are the type and amount of information available about the network. All the strategies (except S1) exploit some information about the network and appropriately assign weights to the links. However, strategies S2 differs from S3 in the type of information required. The EBC based strategy requires global knowledge of the network while the ED based strategy requires only local information. Since there is no unique or straightforward way to extend the definition of ED to the case of directed edges, we try to converge on a definition of ED, for the case of directed networks, in a manner that allows us to identify a relationship between ED and EBC. In doing this and comparing the results of S2 with the results of S3, we investigate the possibility of predicting or obtaining the effects of EBC based strategies by using ED based strategies. 

\begin{equation}
\label{eqn:4}
k_i^{tot}k_j^{tot}=k_i^{in}k_j^{in}+k_i^{in}k_j^{out}+k_i^{out}k_j^{in}+k_i^{out}k_j^{out}
\end{equation}

Given a link $e$, we start by defining ED as the product of total-degrees of the source-node $i$ and target-node $j$. A study of the scatter plots of EBC vs ED doesn't yield any interesting relationship. We then decompose the product of total-degrees into sum of products of in and out degrees of $i$ and $j$ (\ref{eqn:4}). We then study scatter plots of EBC versus each of the terms in (\ref{eqn:4}). While there is no discernible dependence for 3 terms, one of the terms, $k_i^{in}.k_j^{out}$, shows a strong positive correlation with EBC, especially for the case of links that connect hubs. To quantify this dependence, we calculate the Pearson correlation coefficient ($\rho$) of the 2 quantities. The values of $\rho$ for the various networks, under consideration, are given in Table \ref{Tab:CorrCoeff}.


Since it is almost always easier to deal with local information, in terms of both data collection and computational ease, identifying this relationship between the centrality of a link and degrees of the node that it connects becomes important. It also makes it possible to design/analyze strategies based on local information (node-degrees) that have effects similar to strategies based on global information (EBC).

\begin{table*}
\centering
\caption{Summary of the network models and real networks used in the study.}
\label{Tab:NetProp}
\begin{tabular*}{\textwidth}{@{\extracolsep{\fill}} |c| c c c c c |c |c| }
\hline
\multirow{2}{*}{Network} & \multirow{2}{*}{Nodes} & \multirow{2}{*}{Links} & \multirow{2}{*}{WCC} & \multirow{2}{*}{SCC} & \multirow{2}{*}{AIPL} & \multirow{2}{*}{Degree-Correlations} & \multirow{2}{*}{Clustering Coefficient} \\
& & & & & & & \\
\hline
Erd{\"o}s R{\'e}nyi (ER) & 2500 & 12500 & 2500 & 2465 & 0.20 & - & - \\
Scale-Free (SF) & 2500 & 8750 & 2500 & 975 & 0.12 & - & - \\
Assortative ER & 2500 & 12500 & 2500 & 2462 & 0.20 & 0.30 $\pm$ 0.001 & - \\
Assortative SF & 2500 & 8750 & 2500 & 970 & 0.12 & 0.35 $\pm$ 0.002 & - \\
Dissortative ER & 2500 & 12500 & 2500 & 2466 & 0.20 & -0.30 $\pm$ 0.001 & - \\
Dissortative SF & 2500 & 8650 & 2500 & 1080 & 0.14 & -0.40 $\pm$ 0.001 & - \\
Watts-Strogatz & 1000 & 6000 & 1000 & 1000 & 0.18 & - & 0.45 $\pm$ 0.002 \\
Clustered SF & 1000 & 4000 & 1000 & 500 & 0.17 & - & 0.23 $\pm$ 0.016 \\
Austin Road & 7388 & 18956 & 7388 & 7381 & 0.02 & 0.23 $\pm$ 0.012 & 0.009 $\pm$ 0.0009 \\
Protein Interactions & 1706 & 6171 & 1615 & 1492 & 0.19 & -0.13 $\pm$ 0.008 & 0.005 $\pm$ 0.0005 \\
Airport & 3425 & 37594 & 3397 & 3354 & 0.25 & 0.047 $\pm$ 0.0005 & 0.45 $\pm$ 0.018 \\
\hline
\end{tabular*}
\end{table*}

\section{Relevant measures}
\label{sec4}
In the event of loss of a link, the network undergoes various structural and behavioural changes. To gauge the performance of the network and satisfactorily understand these changes, it is necessary to identify a set of relevant measures that can quantify these changes and put them in the right perspective. To this end, we choose a set of reliable measures, that clearly bring out the relationship between the changes occurring in the network and their physical implications.

\subsection{Connected Components}
\label{ss41}
To get a global perspective on the changes occurring in the network during the loss of links, the sizes of the largest strongly and weakly connected components are the metrics of choice. In a directed network, a Strongly Connected Component (SCC) is defined as the set of all nodes for which there exists a  directed path from every node in the set to every other node in the set. A similar set of nodes for which there exists an undirected path between all pairs of nodes is defined as the Weakly Connected Component (WCC). Together, the SCC and WCC 
give rise to a macroscopic organization referred to as the \textit{Bow-Tie} structure \cite{broder2000graph,csermely2013structure}. This is again a subclass in a broader set of core-periphery structures \cite{csermely2013structure} where the core consists of high-degree nodes (hubs) and is very dense and strongly connected while the periphery is comprised mostly of low-degree nodes with sparse connectivity among themselves. 
The study of these connected components provides crucial understanding of the large-scale organization in networks and is also at the heart of percolation studies on networks. 

\begin{figure*}[ht]
\centering
\includegraphics[width=\textwidth]{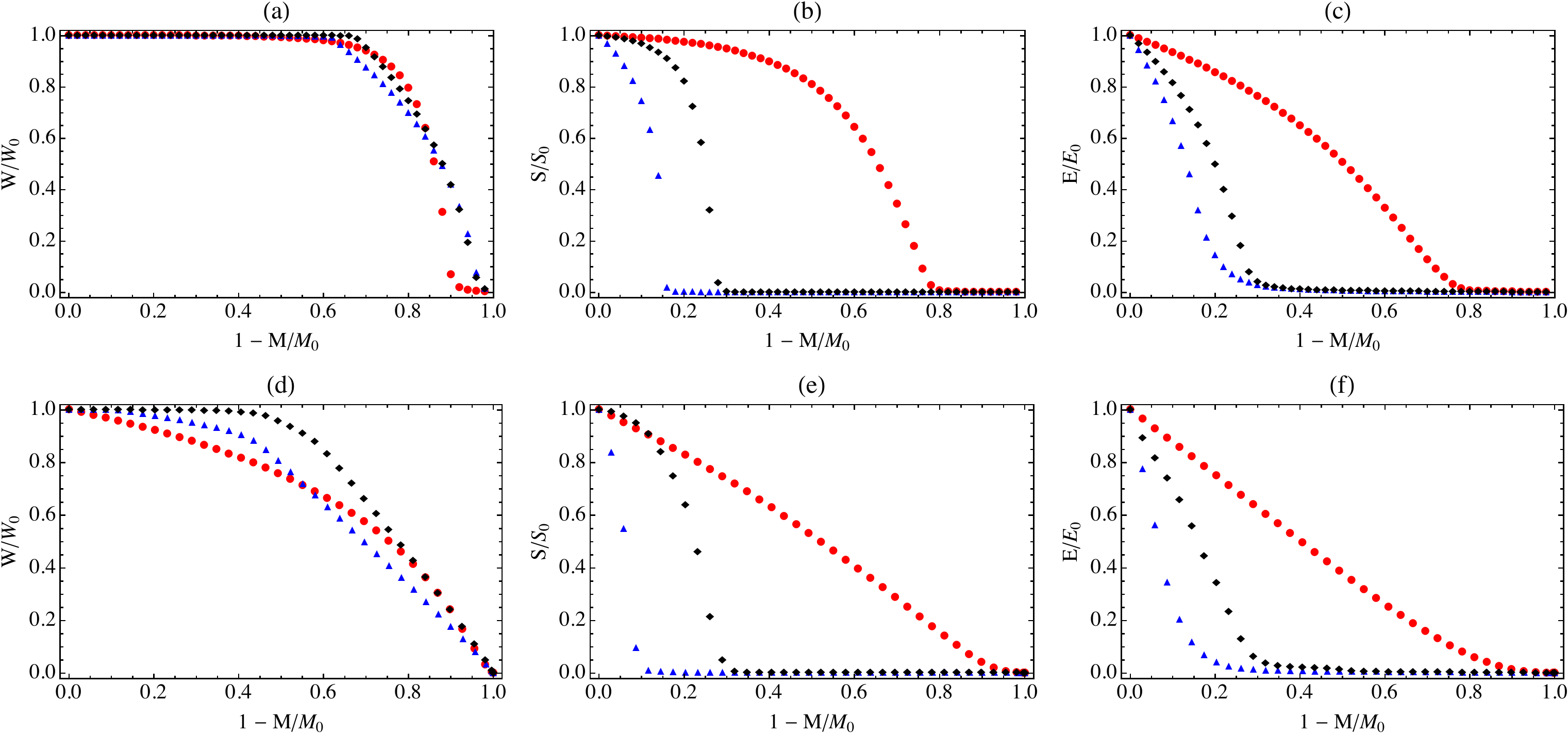}
\caption{(color online) Changes in WCC ($W/W_0$), SCC ($S/S_0$) and Efficiency ($E/E_0$), in ER network (a - c) and SF network (d - f), when they undergo various types of link-deletion. The different strategies of removal are shown as: S1 (red circles), S2 (blue up-triangles), S3 (black diamonds). The measures are plotted against the total fraction of links removed from the network (1 - $M/M_0$).}
\label{fig:Neutral}
\end{figure*}

\subsection{Efficiency}
\label{ss42}
While connected components shed light on structural robustness and the extent of flow/spread in the network, they provide no information on the velocity of a spreading process or information flow in the network. This information is accessed by calculating the Average Inverse Path Length (AIPL), also referred to as Efficiency. It is defined as the average of the inverse geodesic lengths between all pairs of nodes in the network \cite{latora2001efficient}. If $d_{ij}$ is the length of the geodesic path from node $i$ to node $j$, in a network of size N, then the Efficiency, E, is given by




\begin{equation}
\label{eqn:6}
E=\frac{1}{N(N-1)}\sum_i\sum_{j,j \neq i}d_{ij}^{-1}
\end{equation}

Since we are dealing with multiple strategies of deletion on multiple types of networks, for easy comparison of results, all measures are normalized by their respective initial values. Therefore, we are now dealing with quantities $W/W_0$, $S/S_0$ and $E/E_0$, where $W_0$, $S_0$ and $E_0$ are the initial values of largest weakly connected component, largest strongly connected component and efficiency respectively. Equipped with the above measures, to capture the changes occurring in the networks, we now discuss the results of link-deletion in the various networks. 

\section{Results}
\label{sec5}

In this section, we analyze the performance of each type of networks discussed in Sec-II, when they are subject to the various strategies of link-deletion as explained in Sec-III, in terms of the metrics described in Sec-IV. For all network models and the case of random deletion, the measures are calculated as averages over an ensemble of 500 networks. The various measures are plotted against the fraction of total links removed from the network (1 -$M/M_0$). We first study the behavior of simple ER and SF networks under various strategies of deletion. After that, we look at the role of assortative and dissortative degree-correlations on the resilience of these networks. Then, we investigate the contribution of clustering to the same. Finally, we also study the robustness of the three real networks to various kinds of link-deletion and analyze the results.

\subsection*{Link Deletion in ER networks}
In ER networks (Fig.\ref{fig:Neutral} (a - c)), the fractional size of the largest weakly connected component $(W/W_0)$ shows strong robustness to all strategies of deletion. $W/W_0$ remains close to 1 until atleast 60\% of the links are deleted.
The largest strongly connected component ($S$), however,  responds differently to different strategies of removal. During random deletion of links (S1), $S/S_0$ is robust initially and after losing more than half the initial number of links, it begins to breakdown rapidly. For (1-$M/M_0)\geq0.8$, there is no $S$. This value of 0.8 also corresponds to the critical value of $1/N$, for structural phase-transition in random networks. During S2 and S3 types of removal, the network becomes extremely vulnerable and there is immediate deterioration of $S$. In S2 type of deletion, the SCC breaks down completely on losing less than $\sim20\%$ of the most central links, while in S3 type of deletion, a slightly larger number ($\sim$30\%) of links need to be lost before we can observe the same result. 
The change in efficiency ($E/E_0$) follows along the lines of SCC, at least qualitatively , for all types of deletion. During S2 and S3 types of deletion, the behavior of $E/E_0$ and $S/S_0$ is nearly identical, except at the point of transition to complete breakdown.
It is interesting to note that, during S2 and S3, while the SCC undergoes an accelerated breakdown, the rate of decrease of efficiency is quite constant. 

\begin{figure*}[ht]
\centering
\includegraphics[width=\textwidth]{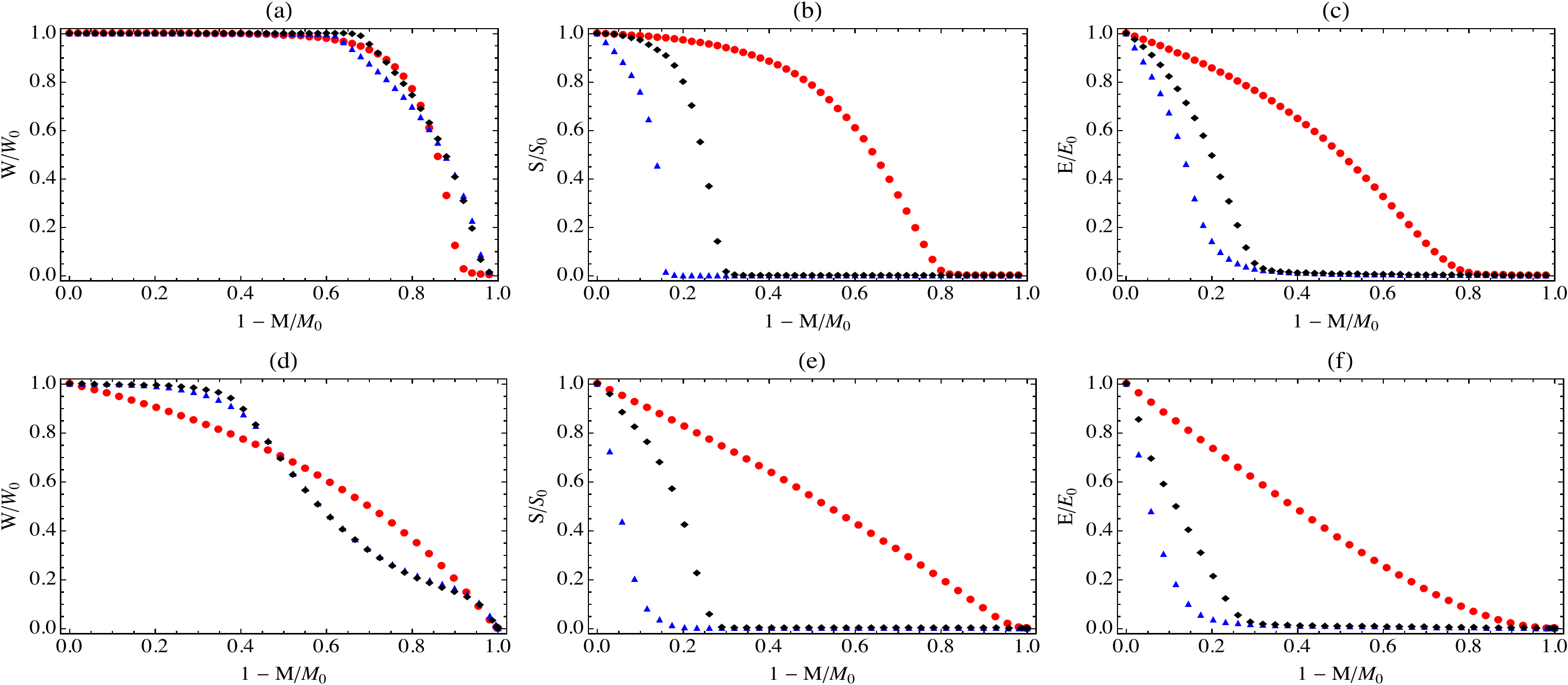}
\caption{(color online) Variation of WCC ($W/W_0$), SCC ($S/S_0$) and Efficiency ($E/E_0$), in assortatively rewired ER networks (a - c) and SF networks (d - f), when they undergo various types of link-deletion.(For details, see Fig.\ref{fig:Neutral}).}
\label{fig:Asrt}
\end{figure*}

\begin{figure*}[ht]
\centering
\includegraphics[width=\textwidth]{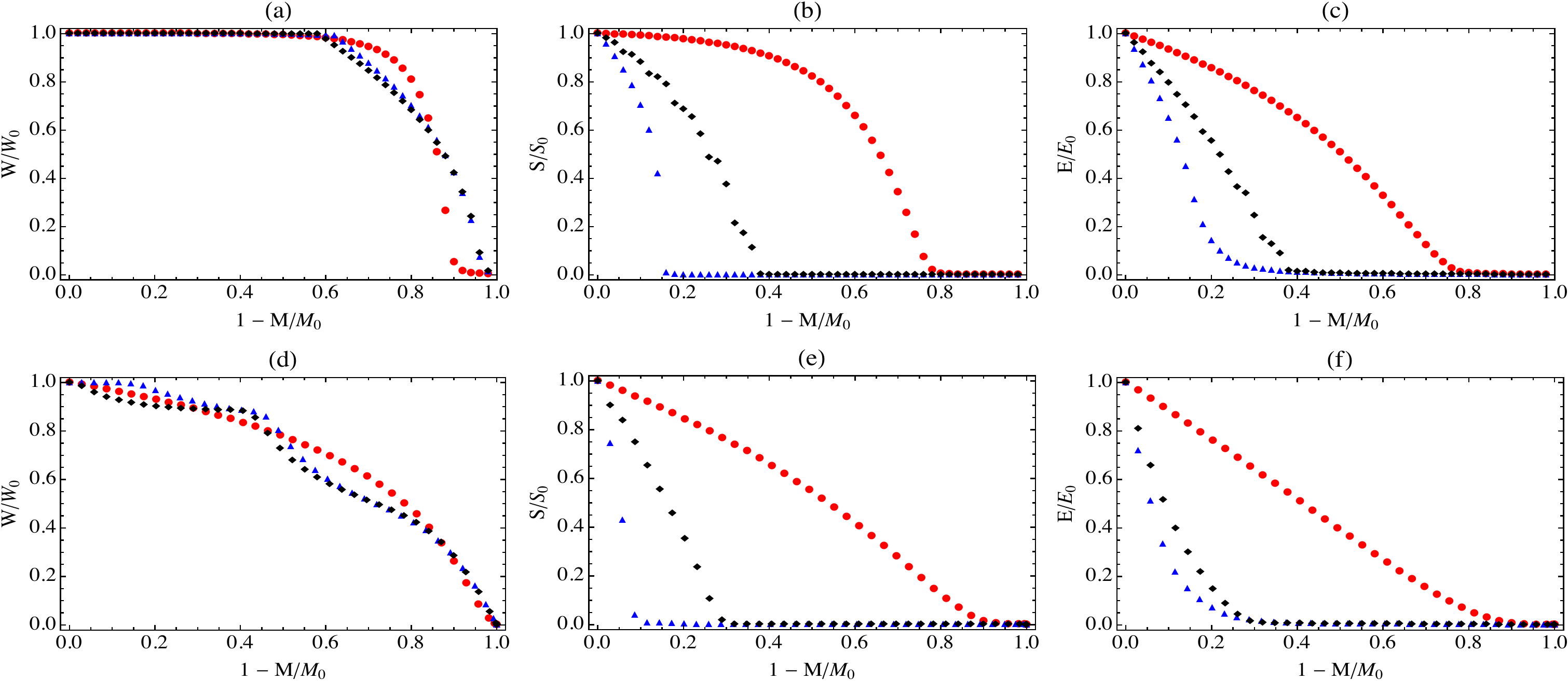}
\caption{(color online) Variation of WCC ($W/W_0$), SCC ($S/S_0$) and Efficiency ($E/E_0$), in dissortatively rewired ER networks (a - c) and SF networks (d - f), when they undergo various types of link-deletion. (Details same as Fig.\ref{fig:Neutral}).}
\label{fig:Dsrt}
\end{figure*}

\subsection*{Link Deletion in SF networks}
In SF networks (Fig.\ref{fig:Neutral} (d - f)), the WCC is again quite robust to all strategies of link deletion. During random deletion, the WCC undergoes a slow and continuous breakdown from the start and becomes more rapid during the later stages of deletion. During S2 and S3, it shows more initial robustness when compared to S1. 
This can be attributed to the core-periphery structure that is emphasized in SF networks. The core contains most of the high-degree nodes and therefore the links within the core have large values of ED. S3 targets these type of links. Due to the high density within the core, the networks stays connected for a while ($1-M/M_0\sim0.4$) before it ultimately breaks down into disconnected components.

When compared to ER networks, strategies S2 and S3 have pretty much the same effect on the SCC in SF networks. S2 leads to complete breakdown of the SCC with just 10\% of the links lost while S3 has the same effect with 30\% loss of links. The response of SCC to S1 is rather different in the case of SF networks. Consistent with the results from percolation studies, during random deletion, the SCC shows a smooth and continuous breakdown without any threshold behavior. 

The behavior of efficiency emphasizes the explanation given for WCC.  During S2 and S3, the efficiency begins to decrease rapidly from the start until $\sim40\%$ of the links are lost. Initially, when the links in the core are being deleted, the network remains largely connected but the connections become progressively less efficient until the nodes get isolated, which corresponds to the sharp decrease in WCC. During S1, there is a continuous decrease with a constant rate.

\subsection*{Role of assortative rewiring in ER and SF networks}
Next, we look at the impact of degree-correlations on the robustness of networks to loss of links. We first study assortatively rewired ER and SF Networks. During assortative rewiring, links are redistributed to connect nodes of similar degrees, while preserving the degree-distributions. As a result, low-degree nodes get connected to other low-degree nodes and hubs connect to other hubs in the network. This makes the core-periphery structure more obvious and effective. The core and the periphery become further well-defined and the interconnections between the two become fewer. The distribution of ED gets skewed and develops an extended tail, indicating the presence of links with very high values. This has direct implications for the strategies based on ED. 

The lack of heterogeneity in the node-degrees in ER networks ensures that the rewiring does not have any substantial effect on the large-scale organization of the network. Therefore, the results of link deletion are similar to the case of neutral ER networks (Fig.\ref{fig:Asrt} (a - c)). 
In SF networks (Fig.\ref{fig:Asrt} (d - f)), where the core-periphery structure is already emphasized, the rewiring process makes it more prominent and the ED-based strategies elicit responses in accordance with this. While the responses to random and EBC-based strategies are not very different from the case of neutral SF network, ED-based strategies bring about a qualitative change in the behavior of WCC and SCC. 
During S3 type of deletion, the WCC shows some initial robustness but not as much compared to the neutral SF network. The initial resilience can be attributed to the links in the highly connected core. Once the links with very high values of ED are removed, there are a smaller number of links with intermediate values (that connect the core and periphery) and the loss of these links leads to the relatively faster collapse seen in the intermediate stages. Towards the end, when the process of deletion reaches the periphery, the links are fewer and the damage more controlled. 

\subsection*{Role of dissortative rewiring in ER and SF networks}
During dissortative rewiring, links are redistributed to connect nodes of dissimilar degrees, in a degree-preserving manner. The distinction between the core and periphery is blurred with a greater fraction of links connecting the core and periphery. The distribution of ED becomes less skewed and is dominated by intermediate values. Once again, in ER networks, the rewiring process does not affect the response to loss of links considerably.
For SF networks (Fig.\ref{fig:Dsrt} (d - f)), the response to random and EBC based deletion is similar to that of neutral SF networks. The effect of smaller concentration of links in the core is clearly visible in the response to ED based strategies. Both, WCC and SCC show reduced resilience when links are removed in decreasing order of ED values (S3). 
The behavior of $E/E_0$ closely follows that of $S/S_0$.

\begin{figure*}[ht]
\centering
\includegraphics[width=\textwidth]{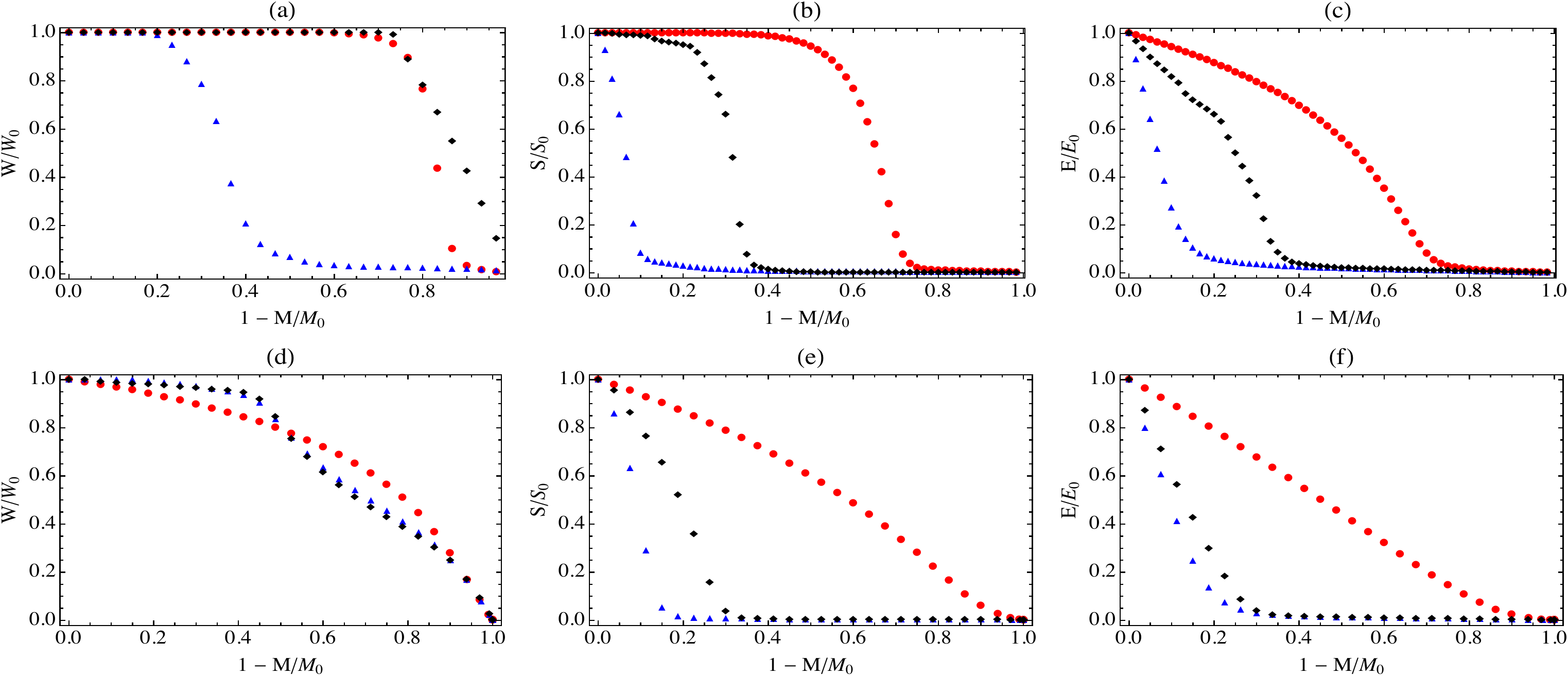}
\caption{(color online) Changes in WCC ($W/W_0$), SCC ($S/S_0$) and Efficiency ($E/E_0$), in directed WS network (a - c) and  clustered SF network (d - f), when they undergo various types of link-deletion. (For details, see Fig.\ref{fig:Neutral}). }
\label{fig:Clust}
\end{figure*}

\subsection*{Role of clustering in ER and SF networks}
To study the independent effect of clustering, on the resilience of networks to loss of links, we use the directed WS model (Fig.\ref{fig:Clust} (a - c)). The generation of this network starts with a 3-regular lattice, following which the links are rewired with a probability $p_r=0.1$. Therefore, on an average, about 10\% of the links are rewired, implying that the network while exhibiting small-world nature, could still retain lattice-like characteristics. In a regular lattice-like structure, there are links with unnaturally high values of betweenness centrality. As the probability of rewiring increases, the network becomes more random, and the values of EBC drop substantially (by an order of magnitude). In our case, since only about 10\% of the links are rewired, the distribution of EBC values has a heavy tail unlike in completely random networks. These edges, with relatively large values of EBC, make the network more susceptible to EBC based strategies, especially S2 type of deletion. Therefore, we see that the SCC become less resilient to S2 type of deletion. With less than 10\% of the links lost, $S/S_0$ shows 90\% breakdown. Even the WCC starts to fall apart after just 20\% of the links have been removed.
In contrast, we observe an increased resilience towards ED based strategies. The response to S3 type of deletion is less effective when compared to ER networks. Both WCC and SCC show relatively higher resistance to loss of links with high values of ED. This could be attributed to the lattice-like structure where there is a considerably large number of alternate paths of comparable lengths. 
In SF networks, 
during random deletion, the results are consistent with those of neutral SF networks. The rewired network is still very susceptible to S2 and S3 strategies. Loss of just 20\% of the links completely destroys the SCC during S2 type of deletion whereas 30\% of the links need to be lost to reach the same stage with S3 type of removal.

\subsection*{Analysis of real-world networks}
\begin{figure*}[ht]
\centering
\includegraphics[width=\textwidth]{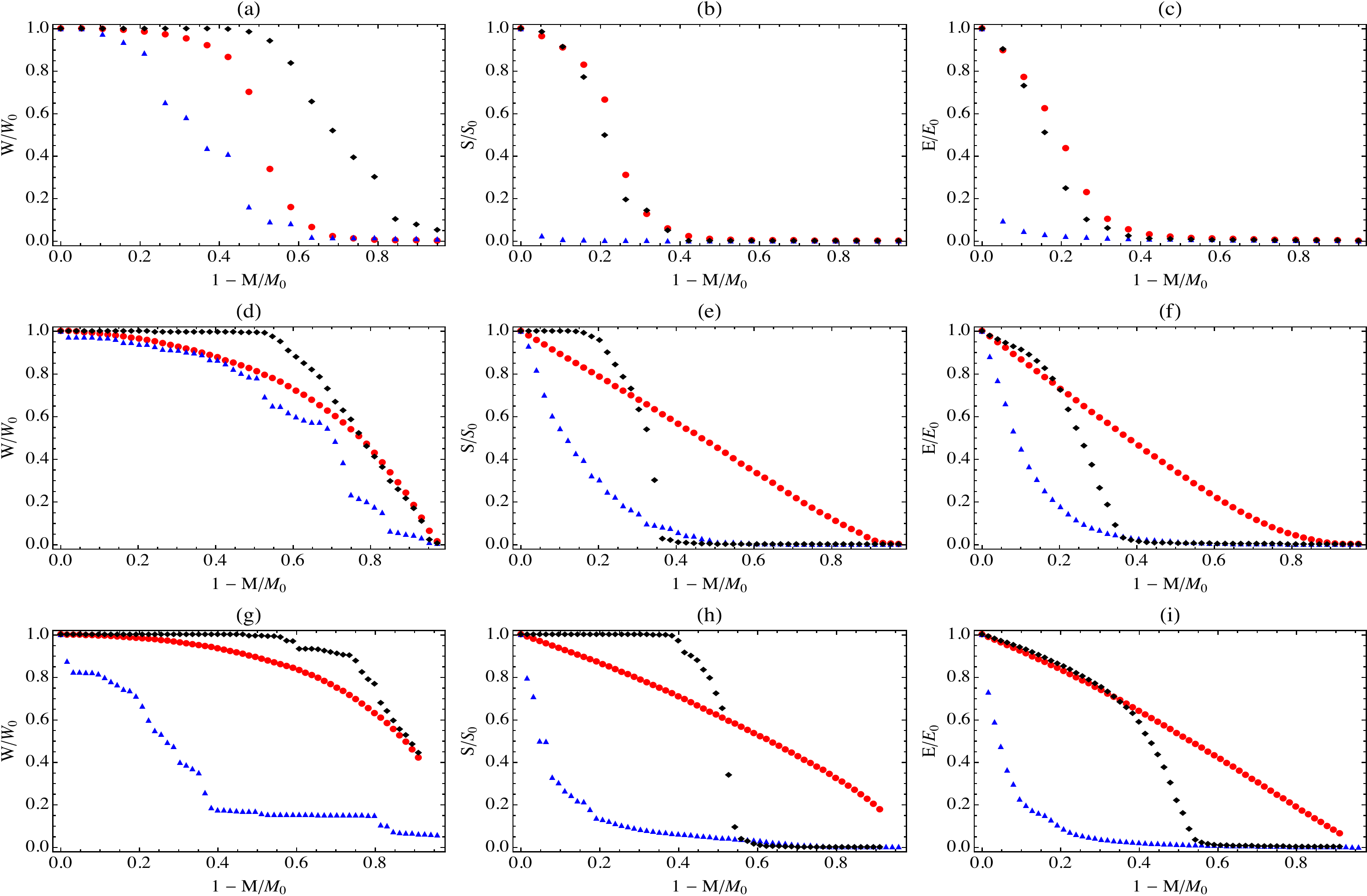}
\caption{(color online) Changes in WCC ($W/W_0$), SCC ($S/S_0$) and Efficiency ($E/E_0$), in the network of road intersections in the city of Austin TX (a - c), the protein interaction network (d - f) and the network of airports across the world (g - i), when they undergo various types of link-deletion. (Details same as Fig.\ref{fig:Neutral}). }
\label{fig:RealN}
\end{figure*}

Next, we study the performance of the three real-networks in the event of targeted or natural link deletion. First, we consider the network of road intersections in the city of Austin TX (Fig.\ref{fig:RealN} (a - c)). The network has no power-law degree distributions and the only conspicuous property is the assortative nature of links. Since it is a transport network, usually, a connection from stop-A to stop-B implies a reciprocal connection from stop-B to stop-A. This need not be true always, like in the case of 1-way lanes or closed-down lanes. Another characteristic feature is that important intersections will have a larger number of roads leading into and out of the intersection and consequently the network has high 1-node correlation.

At first glance, we observe that both connected components and efficiency show little or no resilience to almost all types of link removal. The SCC is extremely vulnerable to S2 becoming non-existent on losing just 10\% of the links but more robust to S1 and S3 types of deletion. Even the WCC does not have any credible resistance to EBC based deletion strategies implying that initial losses could quickly lead to isolated intersections. 

This kind of analysis helps to prepare for any eventuality involving the closure of roads and rerouting of traffic for whatever reason. For example, if certain roads need to be shutdown for maintenance, this analysis helps to select roads so that there is minimum effect of the connectivity and also helps to plan for rate of progress of maintenance. If an alternate scenario, if certain roads are not accessible due to unforeseen circumstances like accidents or traffic-jams, we can predict the loss of further connectivity and traffic pile-ups and take necessary action.

Next, we look at the Protein Interaction Network, which is SF and dominantly dissortative. As would be expected, the response of the network closely resembles that of the dissortatively rewired SF network (Fig.\ref{fig:RealN} (d - f)). During random loss of links, the WCC is slow to disintegrate while the SCC and Efficiency have a steady continuous decline. The WCC is considerably resistant to the EBC based strategy (S2) suggesting a decentralized organization in the network. The SCC, however, is not equally resistant to S2 and undergoes an exponential breakdown. 
The WCC 
is also quite resilient to strategy S4. This suggests that the proteins undergoing fewer interactions play an important role in sustaining the functional connectivity in the proteome. The SCC offers some initial resistance to loss of links with high ED values but undergoes sudden and extensive disintegration soon after. Overall, the network exhibits robustness with respect to the WCC but the SCC is relatively more vulnerable and the efficiency follows along the lines of SCC.

Finally, we present the results of our study on the network of airports across the world (Fig.\ref{fig:RealN} (g - i)). It shows power-law degree distributions for both in and out degrees and has a large value of clustering coefficient. Being a transport network, it shows substantial 1-node correlation but very low values of 2-node degree-correlations. We investigate the effects of link deletion, due to the different strategies, on this network. In doing so, we attempt to validate the results obtained by studying the corresponding network models.

\begin{table*}
\centering
\caption{Summary of results indicating relative robustness of various networks to the different strategies of link deletion and the effectiveness of a given strategy on the different types of networks. The values presented are for the loss of 15\% of the initial number of links in all cases.}
\label{Tab:NetResults}
\begin{tabular*}{\textwidth}{@{\extracolsep{\fill}} |c| c c c |c c c |c c c| }
\hline
\multirow{2}{*}{Network} & \multicolumn{3}{c|}{$W/W_0$} & \multicolumn{3}{c|}{$S/S_0$} & \multicolumn{3}{c|}{$E/E_0$} \\
& S1 & S2 & S3 & S1 & S2 & S3 & S1 & S2 & S3 \\
\hline
Erd{\"o}s R{\'e}nyi (ER) & 0.99 & 1.0 & 1.0 & 0.98 & 0.40 & 0.92 & 0.90 & 0.43 & 0.70 \\
Scale-Free (SF) & 0.94 & 0.99 & 0.99 & 0.874 & 0.004 & 0.82 & 0.81 & 0.10 & 0.47 \\
Assorative ER & 0.99 & 1.0 & 1.0 & 0.98 & 0.33 & 0.92 & 0.90 & 0.40 & 0.7 \\
Assorative SF & 0.93 & 0.99 & 0.99 & 0.87 & 0.01 & 0.66 & 0.85 & 0.04 & 0.38 \\
Dissorative ER & 0.99 & 1.0 & 0.99 & 0.98 & 0.29 & 0.81 & 0.90 & 0.40 & 0.68 \\
Dissorative SF & 0.94 & 0.99 & 0.91 & 0.88 & 0.006 & 0.53 & 0.78 & 0.14 & 0.28 \\
Watts-Strogatz & 1.0 & 1.0 & 1.0 & 0.99 & 0.04 & 0.96 & 0.90 & 0.10 & 0.72 \\
Clustered SF & 0.95 & 0.99 & 0.98 & 0.90 & 0.05 & 0.65 & 0.84 & 0.25 & 0.41 \\
Austin Road & 0.99 & 0.93 & 1.0 & 0.82 & 0.004 & 0.77 & 0.60 & 0.03 & 0.5 \\
Protein Interactions & 0.97 & 0.96 & 1.0 & 0.84 & 0.41 & 0.99 & 0.8 & 0.27 & 0.84 \\
Airport & 0.98 & 0.74 & 1.0 & 0.89 & 0.21 & 1.0 & 0.86 & 0.14 & 0.89 \\
\hline
\end{tabular*}
\end{table*}

Under random loss of links, the WCC and SCC show immediate and continuous reduction in relative sizes. This continuous decomposition of the connected components and the absence of a threshold for breakdown, is consistent with results for directed scale-free networks. The WCC shows strong resilience to strategy S4. The value of $W/W_0$ remains close to 1 until the loss of 80\% of the links. During S2 type of removal, the WCC shows little resilience before undergoing an exponential deterioration. 
The behavior of SCC is similar to that of the WCC. During S4, the SCC shows initial robustness but not as extensively as the WCC. When links with high values of ED are lost, the SCC undergoes a sharp transition after loss of 50\% of the links. When links with high betweenness centrality are removed (S2), there is immediate exponential collapse. 

The vulnerability of the airport network to targeted removal of links connecting low-degree nodes, is most conspicuous from the behavior of efficiency. The physical implication of this observation shows that cancellation of flights between major airports does not affect the extent of connectivity but only makes it less efficient and more cumbersome. The cancellation of flights between smaller airports, for whatever reason, risks isolating them and immediately decreases the extent of connectivity in the overall network. In this network, the flights connecting smaller airports to the bigger hubs play a very central role and loss of such connections leads to immediate and extensive damage, both in terms of extent and efficiency of connectivity. 

\section{Conclusions}
\label{sec6}
We present the results of a comprehensive study of the effects of link deletion in directed networks, with random and scale-free topologies, when subject to various kinds of removal of links, like random failures, EBC and ED based deletions. We further explore the effect of 2-node degree-correlations, on the robustness to loss of links, in both the network topologies. We then investigate the role of clustering, in both the network topologies, in the behavioral response to the various strategies of deletion. The required correlations and clustering are introduced in the networks by means of degree preserving rewiring techniques.

The resilience to loss of links is measured by studying the changes in fractional sizes of the strongly and weakly connected components and the efficiency of the networks. Overall, we find ER networks to be more robust compared to SF networks, irrespective of the strategy of removal. We see that assortative rewiring does not affect the robustness of ER networks but, in SF networks, it decreases the resilience of SCC and WCC to any type of targeted deletion. Dissortative rewiring, in both ER and SF networks, decreases the robustness of SCC while increasing the robustness of WCC. In the directed WS model, presence of clustering, increases the vulnerability irrespective of the type of loss. In SF networks, rewiring for clustering, makes it more vulnerable to S3. In terms of effectiveness of the strategies, we find that removing the most central links has the most damaging effect, independent of topology while removing the least central links has the least damaging effect. Random deletion has a relatively moderate effect compared to the EBC and ED based strategies and the network topology has an effect on how it plays out. Our study on real-world networks, chosen from the broad areas of transportation and biological interactions, have implications on the definition of strategies and on the design of alternate measures to counter or repair the damages of loss of connectivity in them.

We present a summary of the main results in Table \ref{Tab:NetResults}. It gives a comparison of the relative robustness of various networks to different strategies of link deletion.

In an effort to replicate the effectiveness of centrality based strategies, by using just local information about the nodes like degrees, we find that the product of the in and out degrees of the source and target nodes respectively, $k_i^{in}.k_j^{out}$, comes out as an appropriate definition for ED. We find a strong correlation between EBC and ED, particularly for higher values of both, and the same is reflected in the responses that S2 and S3 evoke from the networks. 
This provides a more computationally viable alternative to launch highly effective targeted counterattacks on networks to contain unwanted spreading processes.

It is important to keep in mind that the network models are chosen to represent a broad classification of real-world networks while the choice of strategies is quite arbitrary. The results are meant to present a qualitative picture of the response of network-connectivity to the loss of different types of links. Exact results on real networks can be considerably diverse, depending on their specific properties, and require independent and more focused efforts.




\bibliographystyle{elsarticle-num}
\biboptions{sort&compress}

\end{document}